# Energy Band Gap Engineering of Graphene Nanoribbons

Melinda Y. Han[1], Barbaros Özyilmaz[2], Yuanbo Zhang[2], and Philip Kim[2]
[1]*Department of Applied Physics, Columbia University, New York, New York 10027.*
[2]*Department of Physics, Columbia University, New York, New York 10027.*

**Abstract**

We investigate electronic transport in lithographically patterned graphene ribbon structures where the lateral confinement of charge carriers creates an energy gap near the charge neutrality point. Individual graphene layers are contacted with metal electrodes and patterned into ribbons of varying widths and different crystallographic orientations. The temperature dependent conductance measurements show larger energy gaps opening for narrower ribbons. The sizes of these energy gaps are investigated by measuring the conductance in the non-linear response regime at low temperatures. We find that the energy gap scales inversely with the ribbon width, thus demonstrating the ability to engineer the band gap of graphene nanostructures by lithographic processes.

The recent discovery of graphene [1], a single atomic sheet of graphite, has ignited intense research activities to elucidate the electronic properties of this novel two-dimensional (2D) electronic system. Charge transport in graphene is substantially different from that of conventional 2D electronic systems as a consequence of the linear energy dispersion relation near the charge neutrality point (Dirac point) in the electronic band structure [2, 3]. This unique band structure is fundamentally responsible for the distinct electronic properties of carbon nanotubes (CNTs) [4].

When graphene is patterned into a narrow ribbon, and the carriers are confined to a quasi one-dimensional (1D) system, we expect the opening of an energy gap. Similar to CNTs, this energy gap depends on the width and crystallographic orientation of the graphene nanoribbon (GNR) [5, 6]. However, despite numerous recent theoretical studies [7-15], the energy gap in GNRs has yet to be investigated experimentally.

In this letter, we present electronic transport measurements of lithographically patterned GNR structures where the lateral confinement of charge carriers creates an energy gap. More than two dozen GNRs of different widths and crystallographic orientations were measured. We find that the energy gap depends strongly on the width of the channel for GNRs in the same crystallographic direction, but no systematic crystallographic dependence is observed.

The GNR devices discussed here are fabricated from single sheets of graphene which have been mechanically extracted from bulk graphite crystals onto a $SiO_2$/Si substrate as described in ref. [3]. Graphene sheets with lateral sizes of ~20 μm are contacted with Cr/Au (3/50 nm) metal electrodes. Negative tone e-beam resist, hydrogen silsesquioxane (HSQ), is then spun onto the samples and patterned to form an etch mask defining nanoribbons with widths ranging from 10-100 nm and lengths of 1-2 μm. An oxygen plasma is introduced to etch away the unprotected graphene, leaving the GNR protected beneath the HSQ mask (Fig 1(a)).

In this letter, we study two different types of device sets: device sets P1-P4 each contain many ribbons of varying width running parallel (Fig. 1(b)), and a device sets D1 and D2 have ribbons of uniform width and varying relative orientation (Fig. 1(c)). In either case, each device within a given set is etched from the same sheet of graphene, so that the relative orientation of the GNRs within a given set is known.

We remark that each GNR connects two blocks of wider (~ 0.5 μm) graphene, which are in turn contacted by metal electrodes. Thus, unlike CNTs, Schottky barrier formation by the metal electrodes is absent in our GNR devices. Furthermore, multiple contacts on the wider block of graphene allow for four-terminal measurements in order to eliminate the residual contact resistance (~1 kΩ). A heavily doped silicon substrate below the 300 nm thick $SiO_2$ dielectric layer serves as a gate electrode to tune the carrier density in the GNR. The width ($W$) and the length of each GNR were measured using a scanning electron microscope (SEM) after the transport measurements were performed. Since the HSQ protective layer was not removed from the GNR for this imaging, this measurement provides an upper bound to the true width of the GNR.

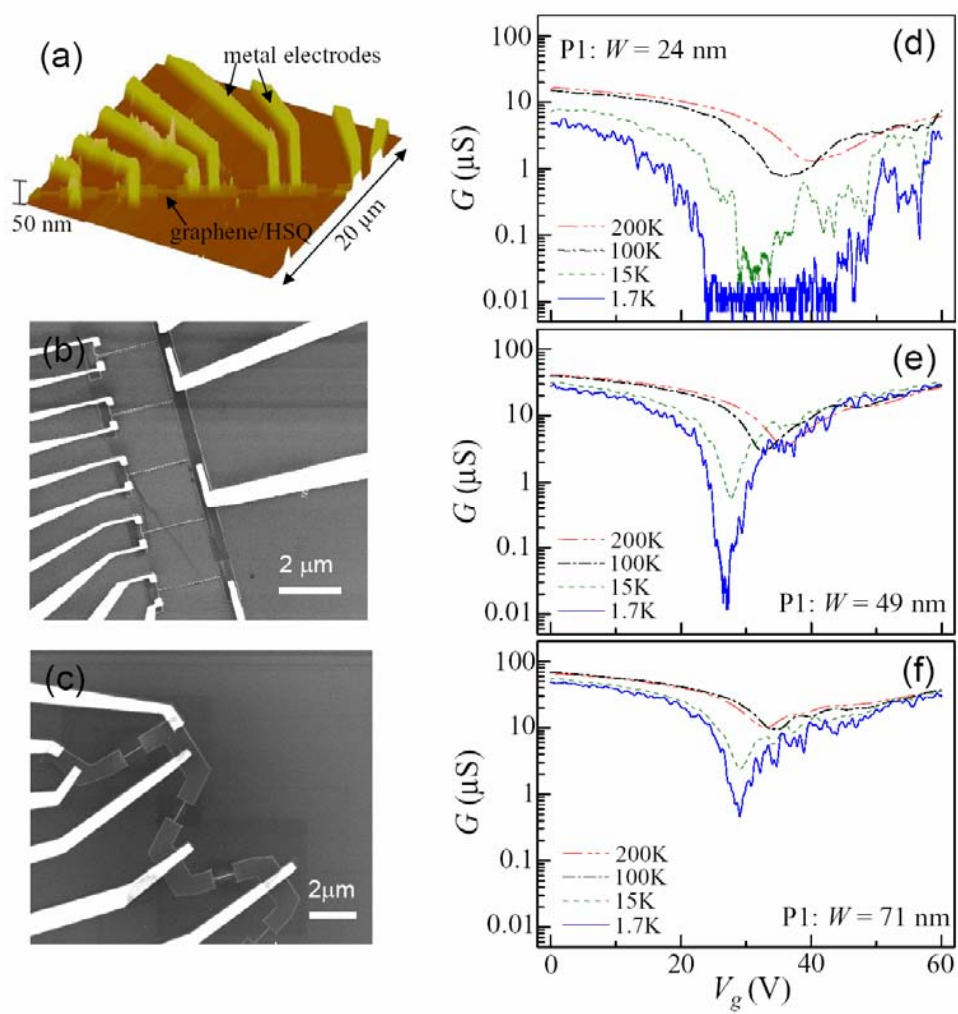

**Figure 1.** (color online) (a) Atomic force microscope image of GNRs in set P3 covered by a protective HSQ etch mask. (b) SEM image of device set P1 with parallel GNRs of varying width. (c) SEM image of device set D2 containing GNRs in different relative crystallographic directions with uniform width (e-f) Conductance of GNRs in device set P1 as a function of gate voltage measured at different temperatures. The width of each GNR is designated in each panel.

The conductance $G$ of the GNRs was measured with a small applied AC voltage (<100uV). Fig. 1(d-f) shows the measured $G$ of three representative GNR devices of varying width ($W = 24 \pm 4, 49 \pm 5$, and $71 \pm 6$ nm) and uniform length ($L = 2$ μm) as a function of gate voltage $V_g$ at different temperatures. All curves exhibit a decrease in $G$ for a range of $V_g$ values. In 'bulk' (i.e., unpatterned) graphene, this dip in $G$ is well understood and corresponds to the minimum conductivity $\sim 4e^2/h$ at the charge neutrality point, $V_g = V_{Dirac}$, where $e$ and $h$ are the electric charge and Plank constant respectively. At room temperature, our GNRs exhibit qualitatively similar $G(V_g)$ behaviors, showing a minimum conductance $G_{min}$ on the order of $4e^2/h\,(W/L)$.

Unlike the bulk case, GNRs with width $W < 100$ nm, show a decrease in $G_{min}$ of more than an order of magnitude at low temperatures. The narrowest GNRs show the greatest suppression of $G_{min}$. For example, for the GNR with $W = 24 \pm 4$ nm (Fig. 1(d)), a large 'gap' region appears for $25 < V_g < 45$ V where $G_{min}$ is below our detection limits ($< 10^{-8} \Omega^{-1}$). This strong temperature dependence of $G(V_g)$ in GNRs is in sharp contrast to that of the 'bulk' graphene samples where $G_{min}$ changes less than 30% in the temperature range 30 mK- 300 K [16]. The suppression of $G$ near the charge neutrality point suggests the opening of an energy gap. We observe (Fig. 1 (d-f)) stronger temperature dependence of $G$ for a broader range of $V_g$ values in narrower GNRs, suggesting larger energy gaps in narrower GNRs.

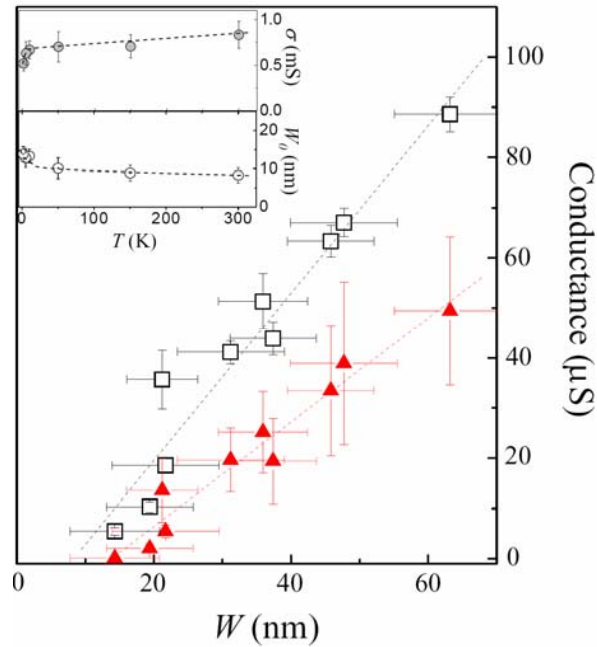

**Figure 2.** (color online) Conductance versus width of parallel GNRs (set P4) measured at $V_g - V_{Dirac} = -50$ V at three representative temperatures. The square and triangle symbols correspond to $T = 300$, and 1.6 K respectively. Dashed lines represent the linear fits at each temperature. The insets show the conductivity (upper) and the inactive GNR width (lower) obtained from the slope and x-intercept of the linear fit at varying temperatures. Dashed curves are shown in the insets as a guide to the eye.

Outside of the 'gap' region near the Dirac point, the conductance scales with the width of the GNR. Fig. 2 shows the conductance of a set of parallel GNRs, with widths ranging from 14-63 nm, measured at two temperatures, $T = 1.6$ and 300 K. The gate voltage is fixed at $V_g = V_{Dirac} - 50$ V, which corresponds to a hole carrier density of $n = 3.6 \times 10^{12}$ cm$^{-2}$. The conductance is well described by the linear fit $G = \sigma(W - W_0)/L$ (dashed line). Here $\sigma$ and $W - W_0$ can be interpreted as the GNR sheet conductivity and the active GNR width participating in charge transport, respectively. The sheet conductivity is ~ 0.75 mS and decreases with decreasing temperature, reaching ~ 75% of the room temperature value at $T = 1.6$ K [17]. The inactive GNR width $W_0$ increases from 10 nm at room temperature to 16 nm at 1.6 K. A reduced active channel width was initially reported in GNRs fabricated on epitaxial multilayer graphene films [18], where much larger inactive edges ($W_0$ ~ 50 nm) were estimated than for our GNR samples. We suggest two possible explanations for the finite $W_0$ measured in our experiment: (i) contribution from localized edge states near the GNR edges due to structural disorder caused by the etching process; (ii) inaccurate width determination due to over-etching underneath the HSQ etch mask. To investigate this, we removed the HSQ etch-mask from several GNRs and found that the actual GNR is often ~ 10 nm narrower than the HSQ protective mask. This suggests that the inactive region due to the localized edge states is small (< 2 nm) at room temperature and spreads to as much as ~ 5 nm at low temperatures. [19]

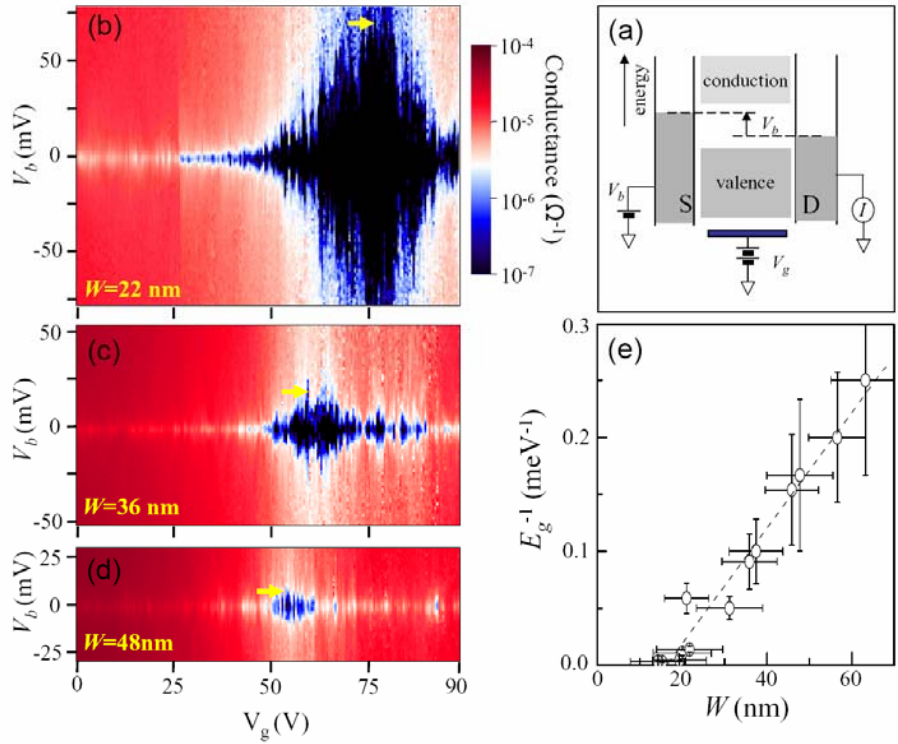

**Figure 3.** (color online) (a) Schematic energy band diagram of a GNR with bias voltage $V_b$ applied. The current $I$ is controlled by both source-drain bias $V_b$ and gate voltage $V_g$. (b-d) The differential conductance ($dI/dV_b$) of three representative GNRs from set P4 with $W$ = 22, 36, and 48 nm as a function of $V_b$ and $V_g$ measured at $T = 1.6$ K. The light (dark) color indicates high (low) conductance as designated by the color map. The horizontal arrows represent $V_b = E_{gap}/e$. (e) $E_{gap}^{-1}$ versus $W$ obtained from similar analysis as (b-d), with a linear fit of the data.

We now discuss the quantitative scaling of the energy gap as a function of GNR width. By examining the differential conductance in the non-linear response regime as a function of both the gate and bias voltage, we can directly measure the size of the energy gap [20]. Fig. 3(a) shows a schematic energy band diagram for a GNR with source and drain electrodes. As the bias voltage, $V_b$, increases, the source and drain levels approach the conduction and valence band edges, respectively. When conduction (valence) band edge falls into the bias window between the source and drain electrodes, electrons (holes) are injected from source (drain) and the current $I$ rises sharply. The gate voltage adjusts the position of the gap relative to the source-drain levels. Fig. 3(b-d) shows the conductance versus $V_g$ and $V_b$ for three representative GNR devices of different width measured at $T = 1.6$ K. The color indicates conductivity on a logarithmic scale, with the large dark area in each graph representing the turned-off region in the $V_g$-$V_b$ plane where the both band edges are outside of the bias windows. The diamond shape of this region indicates that both $V_b$ and $V_g$ adjust the position of the band edges relative to the source and drain energy levels, analogous to non-linear transport in quantum dots [20]. As designated by the arrows, the GNR band gap $E_{gap}$ can be directly obtained from the value of $V_b$ at the vertex of the diamond.

In order to obtain the quantitative scaling of $E_{gap}$ with respect to $W$, we now plot $E_{gap}^{-1}$ against $W$ in Fig. 3(e) for a set of 13 parallel GNRs. The dashed line indicates a linear fit to the data, corresponding to $E_{gap} = \alpha/(W - W^*)$, where we obtained $\alpha = 0.2$ eV/nm and $W^* = 16$ nm from the fit. Recent density functional theory studies [13, 14] predict that the energy gap of a GNR scales inversely with the channel width, with a corresponding $\alpha$ value ranging between 0.2-1.5 eV/nm, which is consistent with this observation. We also note that $W^* \approx W_0$, in good agreement with the independent estimation of the GNR edge effects above.

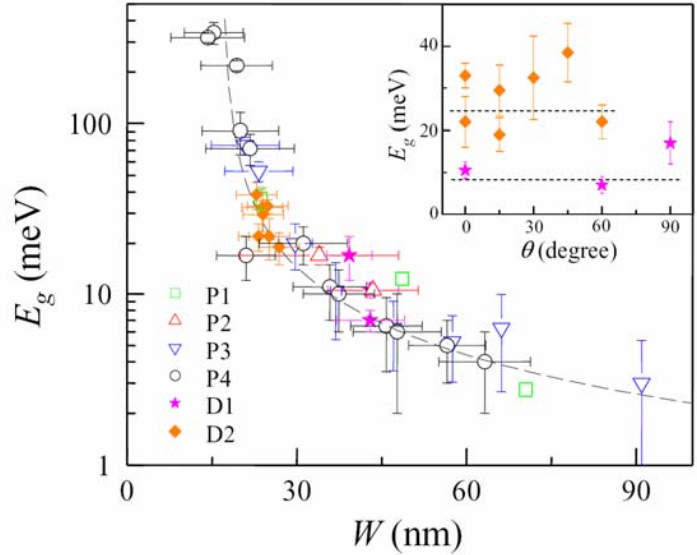

**Figure 4.** (color online) $E_{gap}$ versus $W$ for the 6 device sets considered in this study: four (P1-P4) of the parallel type and two (D1, D2) with varying orientation. The inset shows $E_{gap}$ versus relative angle $\theta$ for the device sets D1 and D2. Dashed lines in the inset show the value of $E_{gap}$ as predicted by the empirical scaling of $E_{gap}$ versus $W$.

A similar scaling behavior holds even across GNR device sets running in different crystallographic directions [21]. Fig. 4 shows the overall scaling of $E_{gap}$ as a function of $W$ for six different device sets. Four device sets (P1-P4) have parallel GNRs with $W$ ranging from 15-90 nm, and two device sets (D1, D2) have GNRs with similar $W$ but different crystallographic directions. The energy gap behavior of all devices is well described by the scaling $E_{gap} = \alpha/(W - W^*)$ as discussed above, indicated by the dashed line. Remarkably, energy gaps as high as ~200 meV are achieved by engineering GNRs as narrow as $W$ ~ 15 nm. Based on the empirical scaling determined here, a narrower GNR may show an even larger band gap, making the use of GNRs for semiconducting device components in ambient conditions a possibility.

Finally, we remark on the crystallographic directional dependence of $E_{gap}$. The inset to Fig 4 shows the variation of $E_{gap}$ for two sets of GNRs with the different relative angle $\theta$. In principle, we expect $E_{gap}(\theta)$ for each set to be periodic in $\theta$, provided all GNRs in the set have similar edge structures. However, experimental observation shows randomly scattered values around the average $E_{gap}$ corresponding to $W$ with no sign of crystallographic directional dependence. This suggests that the detailed edge structure plays a more important role than the crystallographic direction of the GNRs. Indeed, theory for ideal GNRs predicts that $E_{gap}$ depends sensitively on the boundary conditions at the GNR edges [6-15]. At this point, our device fabrication process does not give us atomically precise control of the GNR edges. The interplay between the precise width, edge orientation, edge structure and chemical termination of the edges in GNRs remains a rich area for future research.

In conclusion, we demonstrate that the energy gap in patterned graphene nanoribbons can be tuned during fabrication with the appropriate choice of ribbon width. An understanding of ribbon dimension and orientation as control parameters for the electrical properties of graphene structures can be seen as a first step toward the development of graphene-based electronic devices.

We thank W. de Heer, C. T. White, F. Liu, S. G. Louie, M. Hybertsen, K. Bolotin and P. Jarrillo-Herrero for helpful discussions. This work is supported by the ONR (N000150610138), FENA, NSF CAREER (DMR-0349232) and NSEC (CHE-0117752), and the New York State Office of Science, Technology, and Academic Research (NYSTAR).

*Note added.-* During the preparation of this manuscript we became aware of related work with a similar conclusion from Chen et al. [22].